\documentclass[conference,letterpaper]{IEEEtran}
\usepackage{cite}
\usepackage{amsmath,amssymb,amsfonts}
\usepackage{algorithmic}
\usepackage{graphicx}
\usepackage{textcomp}
\usepackage{xcolor}
\usepackage{hyperref}
\usepackage{siunitx}
\usepackage{svg}
\DeclareSIUnit\sop{SOP}
\DeclareSIUnit\inference{inference}
\DeclareSIUnit\GE{GE}
\DeclareSIUnit\pjsop{\pico\joule\per\sop}
\DeclareSIUnit\njinf{\nano\joule\per\inference}
\DeclareSIUnit\ujinf{\micro\joule\per\inference}
\DeclareSIUnit\GSOPs{\giga\sop\per\second}

\def\BibTeX{{\rm B\kern-.05em{\sc i\kern-.025em b}\kern-.08em
    T\kern-.1667em\lower.7ex\hbox{E}\kern-.125emX}}
\begin{document}

\title{LOKI: a 0.266 pJ/SOP Digital SNN Accelerator with Multi-Cycle Clock-Gated SRAM in 22nm}
\newcommand{\figref}[2]{Fig.~\ref{#1}#2}
\author{\IEEEauthorblockN{Rick Luiken$^\star$, Lorenzo Pes, Manil Dev Gomony and Sander Stuijk}
\IEEEauthorblockA{Dept. of Electrical Engineering, Eindhoven University of Technology, The Netherlands \\
\textit{$^\star$Correspondence to h.j.luiken@tue.nl}
}}

\maketitle

\begin{abstract}
Bio-inspired sensors like Dynamic Vision Sensors (DVS) and silicon cochleas are often combined with Spiking Neural Networks (SNNs), enabling efficient, event-driven processing similar to biological sensory systems. To realize the low-power constraints of the edge, the SNN should run on a hardware architecture that can exploit the sparse nature of the spikes. In this paper, we introduce LOKI, a digital architecture for Fully-Connected (FC) SNNs. By using Multi-Cycle Clock-Gated (MCCG) SRAMs, LOKI can operate at \qty{0.59}{\volt}, while running at a clock frequency of \qty{667}{\MHz}. At full throughput, LOKI only consumes \qty{0.266}{\pjsop}.  We evaluate LOKI on both the Neuromorphic MNIST (N-MNIST) and the Keyword Spotting (KWS) tasks, achieving \qty{98.0}{\percent} accuracy at \qty{119.8}{\njinf} and \qty{93.0}{\percent} accuracy at \qty{546.5}{\njinf} respectively.
\end{abstract}

\section{Introduction}

Bio-inspired event-driven sensors have emerged as a promising solution to address the energy efficiency demands of edge computing. These sensors, inspired by biological systems, transmit information through sparse events known as spikes. Using Address-Event Representation (AER), these spikes are efficiently encoded as asynchronous address events, significantly reducing redundant data transmission. Examples include Dynamic Vision Sensors (DVS) \cite{dvs}, which capture changes in visual scenes with high temporal precision, and silicon cochleas \cite{cochlea}, which emulate the human auditory system by converting sound into spike-based signals. 

Processing the sparse outputs of event-driven sensors demands specialized computational paradigms. Spiking Neural Networks (SNNs), recognized as the third generation of neural networks \cite{maass1997networks}, are particularly well-suited for this task. Unlike traditional Artificial Neural Networks (ANNs), SNNs utilize neurons that process information in the form of discrete spike events, enabling them to handle temporal patterns efficiently. This property makes SNNs particularly advantageous for real-time and low-power applications at the edge.

To harness the potential of SNNs, neuromorphic processors have been developed as dedicated hardware accelerators, optimized for event-driven computation. These processors range from large-scale multi-core architectures like IBM’s TrueNorth \cite{akopyan2015}, Intel’s Loihi \cite{davies2018}, and SpiNNaker 2 \cite{hoppner2022}, to highly energy-efficient single-core designs such as ODIN \cite{frenkel2019}, SNE \cite{dimauro2022}, and ReckOn \cite{frenkel2022}. These single-core designs offer promising solutions for resource-constrained environments.  While these advancements demonstrate significant progress, further energy efficiency improvements remain achievable by adopting advanced memory optimization strategies and high-throughput AER interface techniques.

For instance, advanced memory management techniques, such as Multi-Cycle Clock-Gated (MCCG) SRAM, have emerged as a promising area for power saving \cite{lee2019}. By gating the clock signal to memory cells, MCCG SRAM allows the chip to run at lower voltages while maintaining throughput, thus increasing energy efficiency. Similarly, standard AER protocols are constrained by handshake timing, requiring synchronization with the chip's clock domain. This synchronization can stall the internal pipeline, reducing throughput. In contrast, block AER transmits multiple spikes within a single communication event, significantly enhancing throughput. Despite their potential, these optimization techniques remain unexplored in current neuromorphic accelerators, highlighting an opportunity for further energy optimization.

\begin{figure}[t!]
    \centering   
    \includegraphics[width=\linewidth]{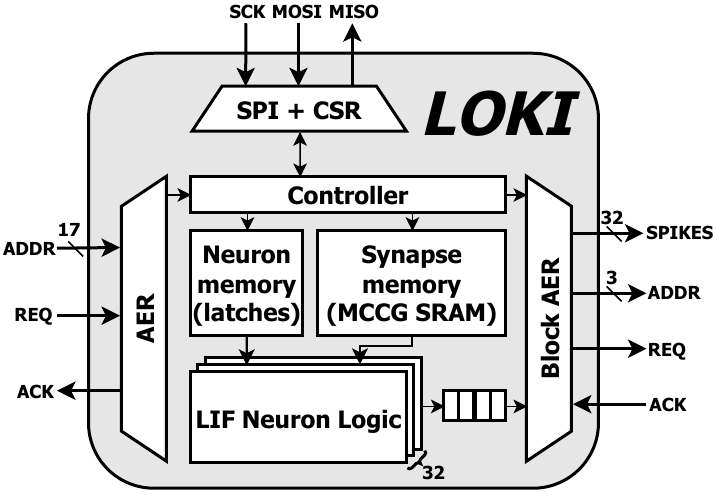}
    \caption{\textbf{Block diagram of the LOKI digital SNN accelerator}. Spikes are received through the AER interface, while parameters are programmed through the SPI interface. Membrane potentials are updated by the LIF Neuron Logic. Output spikes are sent out through the block AER.}
    \label{fig:loki_block_diagram}
\end{figure}

In this work, we present LOKI (\figref{fig:loki_block_diagram}), a novel energy-efficient neuromorphic accelerator designed to address these challenges. The key contributions of this paper are as follows:
\begin{itemize}
    \item We introduce MCCG SRAM for the synapse memory of LOKI. Due to LOKI's regular access pattern, we exploit MCCG SRAM to reduce energy consumption while maintaining throughput compared to standard SRAM. Furthermore, by pipelining multiple spike events, we hide the access latency of MCCG SRAM, increasing the throughput of LOKI.
    \item We introduce the block AER interface, which transmits 32 spikes with one handshake. Block AER significantly improves the interface throughput, preventing stalls in LOKI's pipeline.
    \item We implement LOKI in GF22FDX technology, and evaluate it on two benchmarks: Neuromorphic MNIST (N-MNIST) and Keyword Spotting (KWS). LOKI achieves \qty{98.0}{\percent} accuracy at \qty{119.8}{\njinf} on N-MNIST, and \qty{93.0}{\percent} accuracy at \qty{546.5}{\njinf} on KWS.
\end{itemize}

The remainder of this paper is organized as follows. Section \ref{sec:background} provides relevant background information. Section \ref{sec:architecture} discusses the hardware architecture, and Section \ref{sec:results} presents our experimental results. Finally, Section \ref{sec:conclusion} concludes the paper.

\section{Background}
\label{sec:background}
\subsection{Spiking Neural Networks}

\begin{figure}
    \centering
    \includegraphics[width=\linewidth]{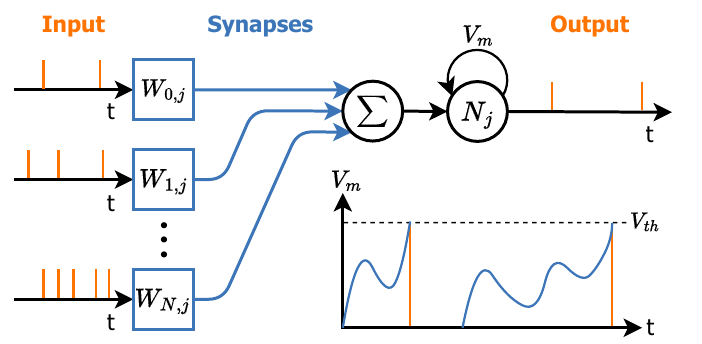}
    \caption{\textbf{SNN behaviour.} Input spikes travel through the synapses, either exciting or inhibiting the neuron's membrane potential ($V_m$). When the membrane potential exceeds the threshold ($V_{th}$), an output spike is emitted.}
    \label{fig:snn}
\end{figure}

SNNs are a biologically inspired class of neural networks where information is processed using discrete spike events, unlike the continuous activations of traditional ANNs. By operating in an event-driven manner as shown in \figref{fig:snn}, SNNs inherently leverage the sparsity of spike-based representations, enabling significant reductions in power consumption and computational overhead. Furthermore, both convolutional and feedforward architectures can be effectively implemented as SNNs, enabling these networks to replicate the feature extraction capabilities of traditional ANNs.

Differently from ANNs, the neuron models employed in SNNs provide a mathematical description of biological neurons, ranging from detailed models like Hodgkin-Huxley (HH) \cite{hh} and Izhikevich \cite{Izhikevich} to simplified approximations such as Leaky Integrate-and-Fire (LIF) \cite{lif}. HH and Izhikevich neurons capture complex spiking behaviors but are computationally demanding. In contrast, the LIF neuron maintains a balance between simplicity and biological plausibility. By abstracting neural activity into a single membrane potential variable, the LIF model reduces complexity while retaining essential features like integration and threshold-based spiking \cite{lif}. This makes it the most widely used neuron model in SNNs, particularly for energy-efficient and event-driven systems. The state and spiking behavior of a LIF neuron are described by the following equations:

\begin{align}
    V_m^t &= \alpha V_m^{t-1} + \sum_jW_{ij}S_j^{t-1} \label{eq:state} , \\
    S_i^t &= H(V_m^t - V_{th}). \label{eq:fire}
\end{align}

\noindent
Here, $V_m^t$ is the membrane potential of the neuron at time $t$, $\alpha$ (where $0.0 \leq \alpha \leq 1.0$) represents the leak factor, and $H$ is the Heaviside step function that outputs a spike $S_i^t$ if the membrane potential exceeds the firing threshold $V_{th}$. The weight matrix $W_{ij}$ is the learnable parameter that governs the synaptic connection strengths between neurons. 

\subsection{Digital SNN accelerators}

SNNs can be efficiently implemented in digital accelerators specialized for low-power execution. These accelerators unlock the full potential of neuromorphic sensors such as DVS and silicon cochleas by bridging the gap between the computational demands of event-based systems and the need for energy-efficient processing platforms.

SNN accelerators are categorized into large-scale multi-core and small-scale architectures. Large-scale platforms, such as Loihi \cite{davies2018} and SpiNNaker 2 \cite{hoppner2022}, prioritize flexibility over energy efficiency for large SNN simulations. In contrast, small-scale architectures like LOKI are designed with energy efficiency in mind. Table \ref{tab:comp} summarizes energy-efficient small-scale architectures. Similar to LOKI, ODIN \cite{frenkel2019}, ESAM \cite{huijbregts2024}, and ANP-I \cite{zhang2023} target Fully-Connected (FC) SNNs, while ReckOn \cite{frenkel2022} targets Recurrent SNNs (SRNNs) and SNE \cite{dimauro2022} targets Convolutional SNNs. ODIN, ReckOn, and ANP-I rely on standard SRAM, lacking memory optimization like MCCG SRAM \cite{lee2019}. SNE reduces memory use via weight sharing across synapses but cannot accelerate FC SNNs. ESAM introduces a compute-in-memory architecture for high throughput but uses binary weights.

\section{Proposed architecture}
In this section, we introduce the architecture of LOKI. We begin by discussing its top-level architecture. Subsequently, we provide detailed descriptions of the subsystem components, including the LIF neuron logic, neuron memory, and block AER interface. Next, we discuss the synapse memory, implemented using MCCG SRAM. We conclude this section by discussing the scalability of the architecture.
\label{sec:architecture}
\subsection{LOKI architecture}

The LOKI architecture is shown in \figref{fig:loki_block_diagram}. LOKI employs a time-multiplexed crossbar with 256 LIF neurons, the membrane potentials of which are stored in the neuron memory. The synapse memory holds $256\times256 = 64$k weights. Membrane potentials are represented as 8-bit signed integers, while the weights are stored as 4-bit signed integers.

The operation of LOKI is as follows: Initially, network parameters, including weights, threshold, and leakage, are written to the chip via SPI. The threshold and leakage parameters are shared across neurons. Once parameters have been written, the chip is ready to receive commands through the AER input interface. During normal operation, LOKI reads spike events from the AER interface until the time reference event is received, marking the end of the current timestep. The time reference event triggers the leak and fire steps of the LIF neurons. First, the membrane potentials of LIF neurons are compared to the threshold, as in (\ref{eq:fire}). If a neuron spikes, its membrane potential is reset to zero. Output spikes are transmitted through the block AER interface. If a neuron does not spike, leakage is applied to its membrane potential. After processing the time reference event, the next timestep begins, allowing new spikes to be processed.

A detailed description of the various modules of LOKI is provided in the following subsection, with the exception of the synapse memory, which is described in Section \ref{sec:MCCG}.

\subsubsection{LIF Neuron Logic} LOKI implements the LIF neuron according to (\ref{eq:state}) and (\ref{eq:fire}). While some accelerators approximate the leakage to be linear \cite{frenkel2019, dimauro2022}, LOKI uses a typical exponential leakage. To enable efficient hardware implementations, we restrict the leakage parameter $\alpha = 1-2^{-k} = 1-\beta$ for $k \in \mathbb{Z}^+$. By restricting the value of the leakage, we can rewrite (\ref{eq:state}) to
\begin{align}
V_m^t &= V_m^{t-1} - \beta V_m^{t-1} + \sum_jW_{ij}S_j^{t-1}. \label{eq:statehardware}
\end{align}
Since dividing by a power of two is the same as shifting to the right, the leakage can be implemented efficiently in hardware. As the membrane potential is signed, leakage always moves the membrane potential towards zero.
\subsubsection{Neuron memory} To update a neuron, we first read its membrane potential from memory, update it using the pre-synaptic weight, and write it back to memory. To increase throughput, the neuron memory is divided into two banks, allowing simultaneous read and write operations. Each bank contains four 256-bit words, each word accounting for the membrane potential of the 32 parallel neurons. Due to the limited size of each bank, the memory is implemented using latches. For compact memories, particularly those with a non-standard aspect ratio of $4 \times 256$, latch-based memories are more area-efficient than standard SRAMs \cite{teman2016power}. 

\subsubsection{Block AER} With the improved throughput provided by LOKI, transmitting output spikes through a standard AER interface becomes a bottleneck. AER employs a 4-phase handshake using request and acknowledge signals, where data is only transmitted when both signals are high. As AER is an asynchronous protocol, it requires the incoming request (for the receiver) or acknowledge (for the sender) to be synchronized within the chip's clock domain. This synchronization is typically achieved using a 2-stage flip-flop synchronizer, which introduces additional delay. This delay, in turn, slows down the 4-phase handshake, reducing the overall transmission speed of the interface. To address this bottleneck, LOKI utilizes a block AER interface for transmitting output spikes. In block AER, a 32-bit spike vector, corresponding to 32 physical neurons in LOKI, is sent along with the 3-bit address of the vector. Consequently, block AER enables the transmission of up to 32 spikes with a single handshake. The original spike addresses are encoded by both the position of each spike within the vector and the accompanying 3-bit address. This approach effectively reduces the bottleneck, preventing stalls in the neuron update pipeline.

\subsection{Synapse Memory with Multi-Cycle Clock-Gated SRAM}
\label{sec:MCCG}
\begin{figure}[t]
    \centering
    \includegraphics[width=0.95\linewidth]{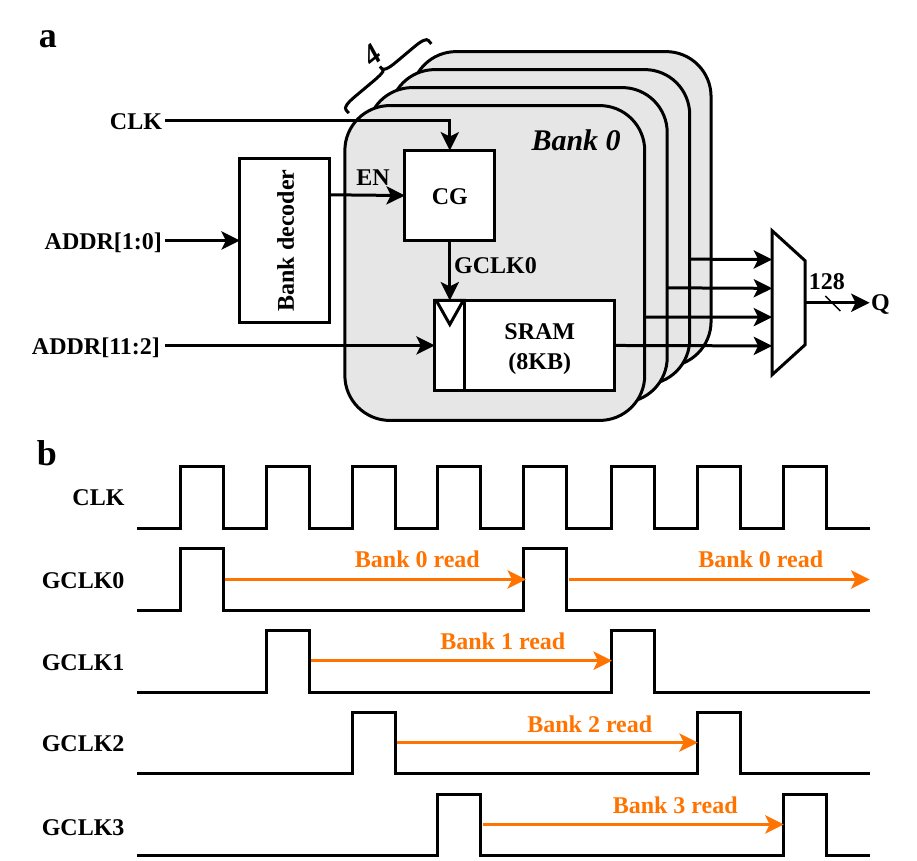}    
    \caption{\textbf{Multi-Cycle Clock-Gated (MCCG) SRAM}. \textbf{a} Block diagram of the synapse memory with MCCG SRAM. The 2 least significant bits of the address are used by the bank decoder to select the appropriate memory bank. The bank decoder then enables the clock for the selected bank, allowing the 9 most significant bits of the address to be captured by the SRAM periphery. The multiplexer at the output selects the bank which most recently completed a read operation. For simplicity, the control logic of the multiplexer is not shown. \textbf{b} Timing diagram of the gated clocks during a sequential read. Each sequential address is stored in a different bank. Each cycle, a different bank is accessed, with each read operation spanning four cycles.}
    \label{fig:MCCG}
\end{figure}

\begin{figure*}[t]
    \centering
    \includegraphics[width=\linewidth]{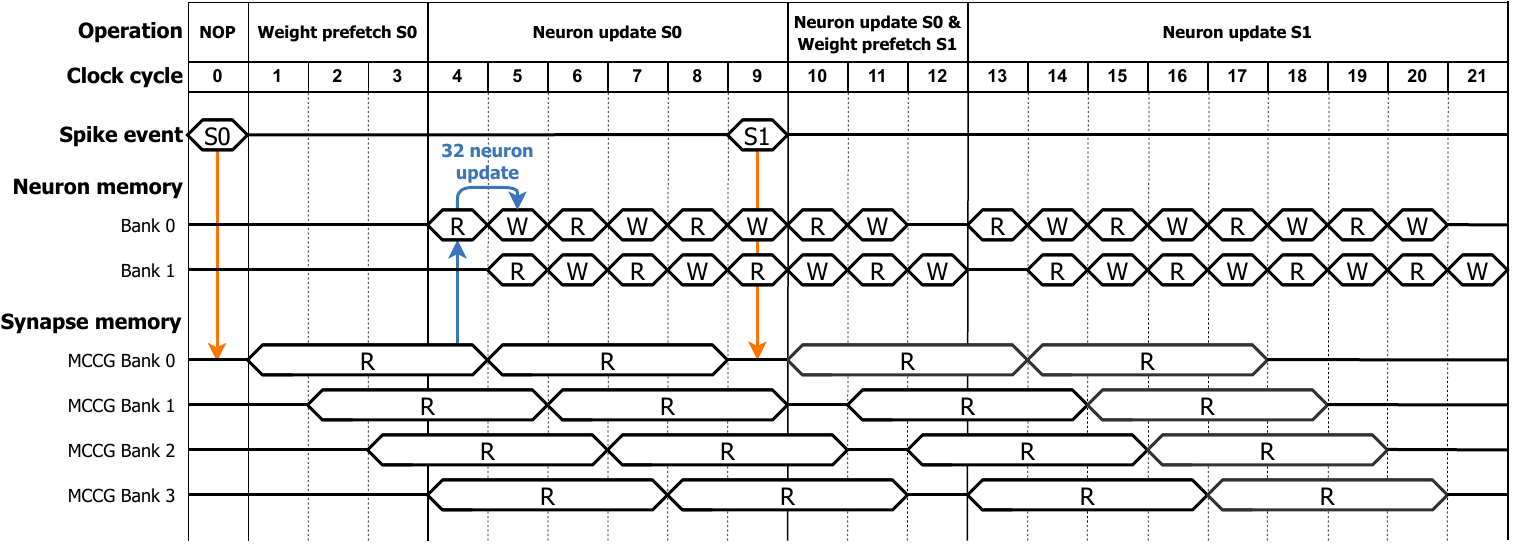}
    \caption{\textbf{Timing diagram of LOKI for two spike events.} After a spike is received through the AER interface, the first post-synaptic weights are fetched. By the 4th clock cycle, the first read operation of MCCG bank 0 is completed. Simultaneously, 32 membrane potentials are read from neuron memory bank 0. These membrane potentials are updated using the weights from MCCG bank 0 and written back into neuron memory bank 0. Within 12 cycles, all 256 neurons are updated. Using pipelining, subsequent spike events are processed in 9 cycles.}
    \label{fig:spiketiming}
\end{figure*}

The synapse memory significantly impacts the energy efficiency of LOKI. For each spike event, 256 weights are read, which requires considerable energy and bandwidth. Although lowering the voltage of the memory reduces energy, it also increases access latency. In designs with a single clock, this latency increase requires the clock frequency to be reduced, significantly reducing the throughput of the accelerator.
To address these issues, we implement the synapse memory using MCCG SRAM, similar to the approach in \cite{lee2019}. The MCCG SRAM, depicted in \figref{fig:MCCG}{a}, consists of four SRAM banks, each implemented using low-voltage commercial SRAMs. These SRAMs consume little energy per access but suffer from high access latency. To avoid slowing down the clock or introducing multiple clock domains, we relax the timing constraints by applying clock gating to the SRAM periphery. \figref{fig:MCCG}{b} shows the timing diagram of the MCCG SRAM clock gating. We enable the clock in each bank once every four cycles, relaxing the timing constraints on each SRAM bank by a factor of four. By reading each bank sequentially, the MCCG SRAM achieves the same bandwidth as a conventional SRAM reading every cycle, while significantly reducing the energy per access due to the lower voltage.

\figref{fig:spiketiming}{} presents a timing diagram of LOKI's operation. It shows the read and write timings for both the neuron memory and the synapse memory, the latter which is implemented using MCCG SRAM. After receiving the first spike event, we begin reading from the synapse memory. Due to the 4-cycle latency of the synapse memory, we first prefetch the post-synaptic weights. By the 4th clock cycle, the first 32 neurons are updated and written back to the neuron memory in the 5th cycle. While bank 0 of the neuron memory is writing back the updated membrane potentials, the next 32 membrane potentials are read from bank 1. Within 12 cycles, all 256 neurons are updated. Additionally, we apply pipelining to increase the throughput, as illustrated in \figref{fig:spiketiming}{}. During cycles 10–12, the weight prefetch for the next spike event occurs concurrently with the neuron updates for the previous event. This overlap reduces the processing time for subsequent spike events to just 9 clock cycles, significantly improving the throughput.

\subsection{Scalability}
In LOKI, we implemented a $256 \times 256$ time-multiplexed crossbar. However, some networks have input layers with more than 256 input synapses per neuron. Additionally, some difficult tasks require more than 256 neurons per hidden layer to achieve the required accuracy. Therefore, to support these use cases, LOKI's design has to be scaled up. Scaling the number of input synapses requires expanding the synapse memory linearly, either by using larger memory banks or by increasing the number of memory banks. Scaling the number of neurons would also require expanding the synapse memory in the same way. Additionally, the neuron memory has to be expanded to store the increased number of membrane potentials. Depending on the number of extra neurons, it could be beneficial to implement the neuron memory using standard SRAM instead of latch-based memory.

\section{Experimental results}
\label{sec:results}
In this section, we evaluate the architecture through synthesis and simulation. The architecture is synthesized in GF22FDX using Cadence Genus 22.1. We used 8T, LVT and SLVT cells, SSG corner, \qty{0.59}{\volt} nominal supply voltage, \qty{-40}{\celsius}. We used a target frequency of \qty{667}{\mega\hertz}. The power consumption is estimated using Genus at TT corner, \qty{0.59}{\volt} supply voltage, \qty{125}{\celsius}. For each experiment, we run an SDF-annotated Gate Level Simulation (GLS) using Cadence XCelium 23.09 and extract the switching activity, which is used in Genus to get the average power consumption during the simulation. For both energy efficiency and throughput, we count the number of Synaptic Operations (SOPs) performed. One SOP is equal to one update to the membrane potential due to an incoming spike. With a layer of 256 neurons, this results in 256 SOPs per incoming spike. We calculate the energy efficiency by dividing the total chip energy by the number of SOPs performed. For the throughput, we divide the number of SOPs performed by the simulation time, excluding the time taken for writing the weights and parameters. This methodology ensures an accurate analysis of the architecture’s energy efficiency and computational performance, as explored in the following subsections, which examine key metrics such as area, energy efficiency, and throughput, as well as practical performance in conventional edge-AI benchmarks for SNNs.

\subsection{Architecture metrics}
\label{sec:architecture_metrics}

\subsubsection{Area} We report the estimated area in gate equivalent (GE) by taking the area estimate provided by the synthesis tool and dividing it by the area of an ND2X1 NAND gate. For LOKI, this results in an estimated area of \qty{337.5}{\kilo\GE}.

\begin{figure*}[t]
    \centering
    \includegraphics[width=0.85\linewidth]{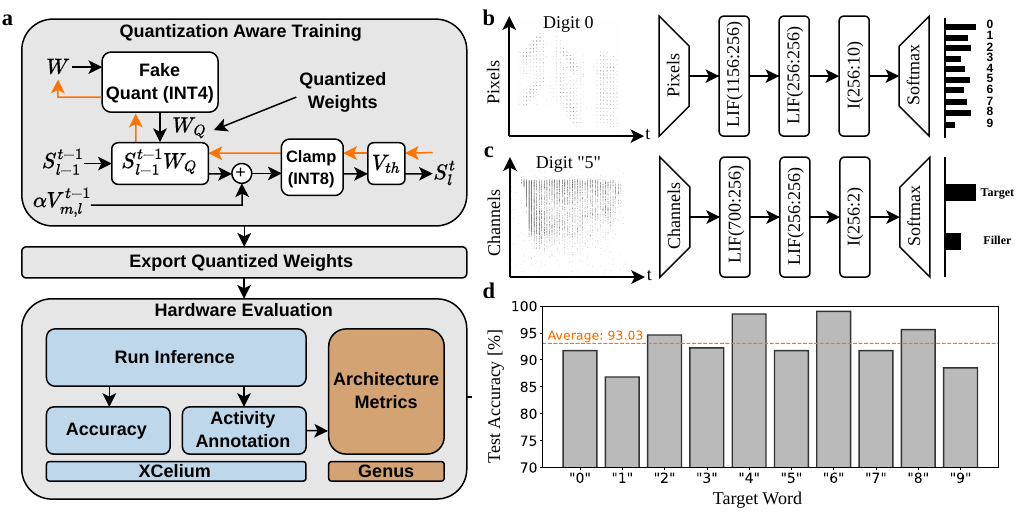}
    \caption{\textbf{Evaluation on Use Cases.} \textbf{a} Evaluation pipeline: the computational graph of QAT is shown at the top, with orange arrows indicating error backpropagation. Gradients update the FP32 weights ($W$). Below, the hardware evaluation process uses quantized weights ($W_{Q}$) in gate-level simulations with XCelium to determine inference accuracy and activity annotations. These annotations are used in Genus to estimate architectural metrics. \textbf{b} Network model for the N-MNIST dataset. \textbf{c} Network model for the 1-word KWS dataset. \textbf{d} Test accuracy for different target words for the 1-word KWS dataset.}
    \label{fig:kws_results}
\end{figure*}

\begin{figure}[t!]
    \centering   
    \includegraphics[width=0.9\linewidth]{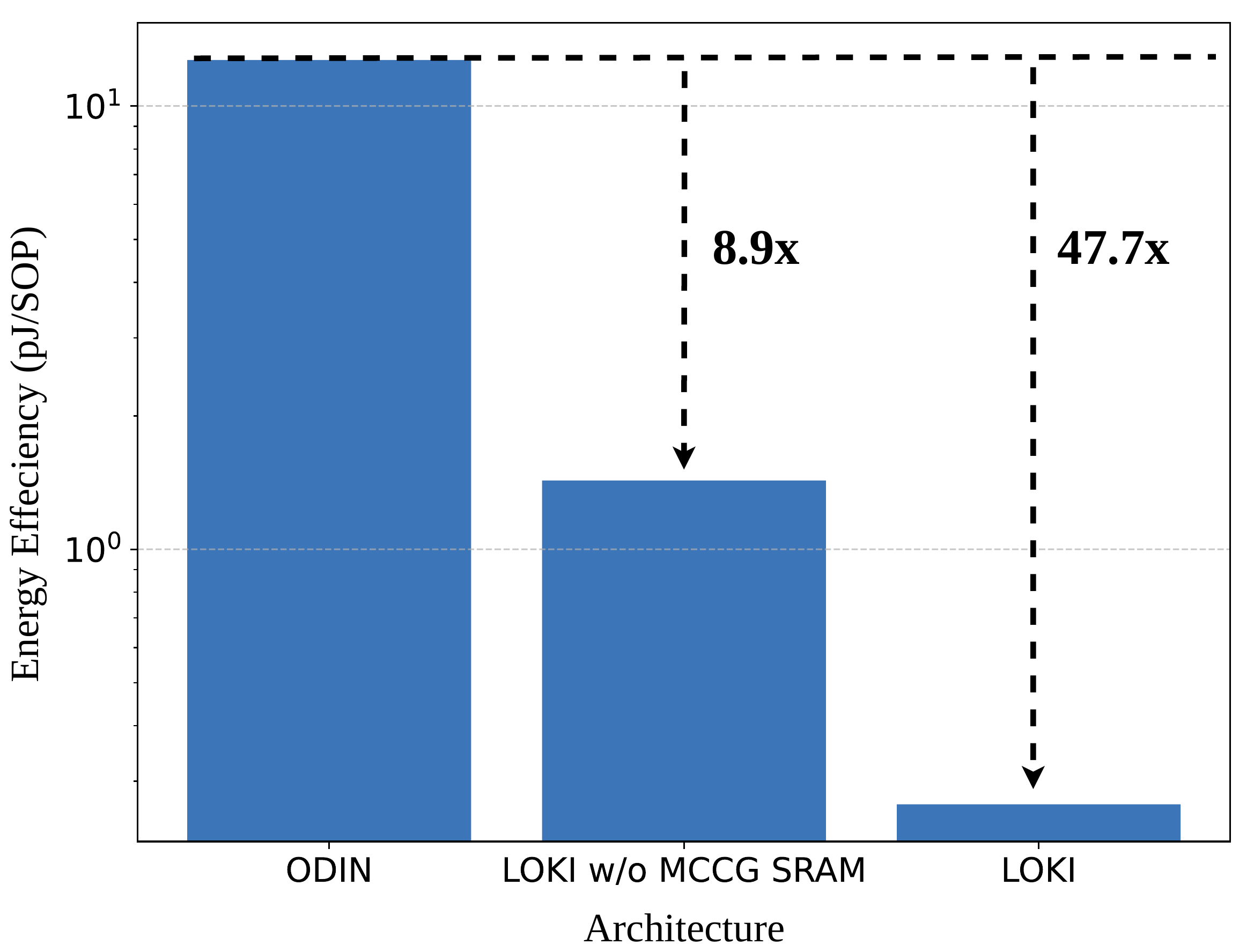}
    \caption{\textbf{Energy Efficiency Gains of LOKI}. Comparison of the energy efficiency of LOKI without MCCG SRAM and LOKI with MCCG SRAM relative to ODIN, highlighting the impact of memory optimization on power consumption.}
    \label{fig:ef_gains}
\end{figure}

\subsubsection{Energy Efficiency and Throughput} To get the peak energy efficiency, we simulate one 256-256 layer with \qty{0}{\percent} input sparsity for 10 timesteps. By having \qty{0}{\percent} input sparsity, we ensure the neuron update pipeline is always filled, which gives the peak utilization. At \qty{0}{\percent} input sparsity, LOKI consumes only \qty{0.266}{\pjsop}. \figref{fig:ef_gains}{} shows the gains in energy efficiency of LOKI compared to two neuromorphic architectures. LOKI is $47.7\times$ more energy-efficient than ODIN, the most similar architecture in the related works. Comparing the MCCG SRAM with standard SRAM, the improvement in energy efficiency is $5.4\times$. The maximum throughput of LOKI is \qty{18.8}{\GSOPs}. By increasing the clock frequency to \qty{667}{\mega\hertz}, updating 32 neurons in parallel and pipelining, LOKI's throughput is $501\times$ higher than ODIN.

\begin{table*}[ht]
\centering
\renewcommand{\arraystretch}{1.2}
\caption{Comparison of neuromorphic accelerators}
\label{tab:comp}
\resizebox{\textwidth}{!}{%
\begin{tabular}{ccccccc}
\hline
\multicolumn{1}{|c|}{\textbf{}} &
  \multicolumn{1}{c|}{\begin{tabular}[c]{@{}c@{}}ODIN \cite{frenkel2019}\\ TBioCAS, 2019\end{tabular}} &
  \multicolumn{1}{c|}{\begin{tabular}[c]{@{}c@{}}ReckOn \cite{frenkel2022}\\ ISSCC, 2022\end{tabular}} &
  \multicolumn{1}{c|}{\begin{tabular}[c]{@{}c@{}}SNE \cite{dimauro2022}\\ DATE, 2022\end{tabular}} &
  \multicolumn{1}{c|}{\begin{tabular}[c]{@{}c@{}}ESAM \cite{huijbregts2024}\\ DAC, 2024\end{tabular}} &
  \multicolumn{1}{c|}{\begin{tabular}[c]{@{}c@{}}ANP-I \cite{zhang2023}\\ ISSCC, 2023\end{tabular}} &
  \multicolumn{1}{c|}{\textbf{\begin{tabular}[c]{@{}c@{}}LOKI\\ This work\end{tabular}}} \\ \hline
\multicolumn{1}{|c|}{\textbf{Technology}} &
  \multicolumn{1}{c|}{\qty{28}{\nm}} &
  \multicolumn{1}{c|}{\qty{28}{\nm}} &
  \multicolumn{1}{c|}{\qty{22}{\nm}} &
  \multicolumn{1}{c|}{\qty{3}{\nm}} &
  \multicolumn{1}{c|}{\qty{28}{\nm}} &
  \multicolumn{1}{c|}{\textbf{\qty{22}{nm}}} \\ \hline
\multicolumn{1}{|c|}{\textbf{Voltage}} &
  \multicolumn{1}{c|}{\qty{0.55-1.0}{\volt}} &
  \multicolumn{1}{c|}{\qty{0.5-0.8}{\volt}} &
  \multicolumn{1}{c|}{\qty{0.8}{\volt}} &
  \multicolumn{1}{c|}{\qty{0.7}{\volt}} &
  \multicolumn{1}{c|}{\qty{0.56-0.9}{\volt}} &
  \multicolumn{1}{c|}{\textbf{\qty{0.59}{\volt}}} \\ \hline
\multicolumn{1}{|c|}{\textbf{Clock frequency}} &
  \multicolumn{1}{c|}{\qty{75-100}{\mega\hertz}} &
  \multicolumn{1}{c|}{\qty{13-115}{\mega\hertz}} &
  \multicolumn{1}{c|}{\qty{400}{\mega\hertz}} &
  \multicolumn{1}{c|}{\qty{810}{\mega\hertz}} &
  \multicolumn{1}{c|}{Async} &
  \multicolumn{1}{c|}{\textbf{\qty{667}{\mega\hertz}}} \\ \hline
\multicolumn{1}{|c|}{\textbf{Network type}} &
  \multicolumn{1}{c|}{FC SNN} &
  \multicolumn{1}{c|}{SRNN} &
  \multicolumn{1}{c|}{Conv SNN} &
  \multicolumn{1}{c|}{FC SNN} &
  \multicolumn{1}{c|}{FC SNN} &
  \multicolumn{1}{c|}{\textbf{FC SNN}} \\ \hline
\multicolumn{1}{|c|}{\textbf{Online learning}} &
  \multicolumn{1}{c|}{SDSP} &
  \multicolumn{1}{c|}{E-Prop} &
  \multicolumn{1}{c|}{-} &
  \multicolumn{1}{c|}{-} &
  \multicolumn{1}{c|}{S-TP} &
  \multicolumn{1}{c|}{-} \\ \hline
\multicolumn{1}{|c|}{\textbf{State/Weight resolution}} &
  \multicolumn{1}{c|}{INT8/INT4} &
  \multicolumn{1}{c|}{INT16/INT8} &
  \multicolumn{1}{c|}{INT8/INT4} &
  \multicolumn{1}{c|}{(N/A)/INT1} &
  \multicolumn{1}{c|}{(N/A)/INT8} &
  \multicolumn{1}{c|}{\textbf{INT8/INT4}} \\ \hline
\multicolumn{1}{|c|}{\textbf{Neuron model}} &
  \multicolumn{1}{c|}{LIF} &
  \multicolumn{1}{c|}{LIF} &
  \multicolumn{1}{c|}{LIF} &
  \multicolumn{1}{c|}{IF} &
  \multicolumn{1}{c|}{LIF} &
  \multicolumn{1}{c|}{\textbf{LIF}} \\ \hline
\multicolumn{1}{|c|}{\textbf{Dataset}} &
  \multicolumn{1}{c|}{MNIST} &
  \multicolumn{1}{c|}{\begin{tabular}[c]{@{}c@{}}IBM DVS Gestures\\ KWS on SHD\\ Delayed Cue Integration\end{tabular}} &
  \multicolumn{1}{c|}{\begin{tabular}[c]{@{}c@{}}IBM DVS Gestures\\ N-MNIST\end{tabular}} &
  \multicolumn{1}{c|}{MNIST} &
  \multicolumn{1}{c|}{\begin{tabular}[c]{@{}c@{}}IBM DVS Gestures\\ N-MNIST\\ KWS on N-TIDIGIT\\ SeNic\end{tabular}} &
  \multicolumn{1}{c|}{\textbf{\begin{tabular}[c]{@{}c@{}}N-MNIST\\ KWS on SHD\end{tabular}}} \\ \hline
\multicolumn{1}{|c|}{\textbf{Accuracy}} &
  \multicolumn{1}{c|}{\begin{tabular}[c]{@{}c@{}}Rate coding: \qty{91.9}{\percent}\\ Rank coding: \qty{91.4}{\percent}\end{tabular}} &
  \multicolumn{1}{c|}{\begin{tabular}[c]{@{}c@{}}Gest: \qty{87.3}{\percent}@10 classes\\ SHD: \qty{90.7}{\percent}@1word\\ DCI: \qty{96.4}{\percent}@2 decisions\end{tabular}} &
  \multicolumn{1}{c|}{\begin{tabular}[c]{@{}c@{}}Gest: \qty{92.8}{\percent}\\ N-MNIST: \qty{97.9}{\percent}\end{tabular}} &
  \multicolumn{1}{c|}{\qty{97.6}{\percent}} &
  \multicolumn{1}{c|}{\begin{tabular}[c]{@{}c@{}}Gest: \qty{92.0}{\percent}\\ N-MNIST: \qty{96.0}{\percent}\\ N-TIDIGIT: \qty{92.6}{\percent}@1word\\ SeNic: \qty{95.7}{\percent}@7classes\end{tabular}} &
  \multicolumn{1}{c|}{\textbf{\begin{tabular}[c]{@{}c@{}}N-MNIST: \qty{98.0}{\percent}\\ SHD: \qty{93.0}{\percent}@1word\end{tabular}}} \\ \hline
\multicolumn{1}{|c|}{\textbf{Energy per inference}} &
  \multicolumn{1}{c|}{\begin{tabular}[c]{@{}c@{}}Rate coding: \qty{451}{\nano\joule}\\ Rank coding: \qty{15}{\nano\joule}\end{tabular}} &
  \multicolumn{1}{c|}{\begin{tabular}[c]{@{}c@{}}Gest: \qty{46}{\micro\joule}\\ SHD: \qty{4.4}{\micro\joule}\\ Nav: \qty{1.3}{\micro\joule}\end{tabular}} &
  \multicolumn{1}{c|}{\begin{tabular}[c]{@{}c@{}}Gest: \qty{80-261}{\micro\joule}\\ N-MNIST: \qty{43-142}{\micro\joule}\end{tabular}} &
  \multicolumn{1}{c|}{\qty{0.6}{\nano\joule}} &
  \multicolumn{1}{c|}{\begin{tabular}[c]{@{}c@{}}Gest: \qty{3.9}{\micro\joule}\\ N-MNIST: \qty{343}{\nano\joule}\\ N-TIDIGIT: \qty{6.1}{\micro\joule}\\ SeNic: \qty{582}{\nano\joule}\end{tabular}} &
  \multicolumn{1}{c|}{\textbf{\begin{tabular}[c]{@{}c@{}}N-MNIST: \qty{28.8/119.8^*}{\nano\joule}\\ SHD: \qty{72.1/546.5^*}{\nano\joule}\end{tabular}}} \\ \hline
\multicolumn{1}{|c|}{\textbf{Energy Efficiency}} &
  \multicolumn{1}{c|}{12.7 pJ/SOP} &
  \multicolumn{1}{c|}{5.3-12.8 pJ/SOP} &
  \multicolumn{1}{c|}{0.221 pJ/SOP} &
  \multicolumn{1}{c|}{N/A} &
  \multicolumn{1}{c|}{1.5 pJ/SOP} &
  \multicolumn{1}{c|}{\textbf{0.266 pJ/SOP}} \\ \hline
\multicolumn{7}{l}{$^*$Estimated total energy for the first two layers of the network. The first layer energy is estimated by multiplying the \si{\pjsop} of the second layer and the number of SOPs performed in the first layer.}
\end{tabular}%
}
\end{table*}
\subsection{Evaluation on Use Cases}

We evaluate the performance of LOKI on two widely used datasets for benchmarking edge-AI SNN accelerators:

\begin{itemize}
    \item \textit{Neuromorphic-MNIST (N-MNIST) \cite{nmnist}}: 
    N-MNIST is a neuromorphic version of the standard MNIST dataset, where handwritten digit images are converted into spike trains using a dynamic vision sensor (DVS). The goal is to classify digits (0–9) based on the spiking activity generated by the DVS. This dataset is particularly well-suited for demonstrating the feasibility of integrating event-based DVS sensors with low-power inference systems, showcasing their potential for efficient operation in energy-constrained edge-AI scenarios.

    \item \textit{1-word Keyword Spotting (KWS)}: 
    The KWS task involves detecting the presence of a specific keyword within an audio stream. We use the Spiking Heidelberg Digits (SHD) dataset \cite{shd} as the source of both target keywords and filler words. Specifically, English digits are used as target keywords, while filler words include a mix of English and German digits. We employ a 1:1 ratio of target-to-filler words to ensure a balanced evaluation.
\end{itemize}

 The evaluation pipeline is depicted in Fig.~\ref{fig:kws_results}{a} while the network architectures employed for N-MNIST and KWS are depicted in Fig.~\ref{fig:kws_results}{b} and Fig.~\ref{fig:kws_results}{c}, respectively. For both models, hidden layers use LIF models with learnable thresholds to adapt firing behavior during training. The output layers are Integrators with a Softmax function applied to the membrane potentials, generating a probability distribution over classes. 

All networks are trained in PyTorch using quantization-aware training (QAT), as illustrated at the top of Fig.~\ref{fig:kws_results}{a}, to accommodate the resolution constraints of the hardware device. Specifically, synaptic connections are quantized to 4-bit signed integers using fake quantization \cite{fake_quant}, and membrane potentials are clamped to 8-bit signed integers to prevent overflow. Notably, gradient updates are applied to the full-precision (FP32) synaptic connections, which are then quantized at each forward pass.  To ensure compatibility with discrete-time simulation frameworks, spike events from both datasets are converted into frame-based inputs using a 10,000 ms time window. Hyperparameter tuning is conducted for layer leakages, learning rates, and batch sizes, with training performed using the Adam optimizer over 20 epochs. To account for the non-differentiable nature of the spiking activation function, we employ the "arctan" surrogate gradient from \cite{surrogate}.

Following QAT, the quantized parameters are deployed on the simulated hardware to evaluate their test accuracy and energy efficiency under realistic operating conditions. The energy metric for specific use cases is calculated using the same procedure detailed in the beginning of this section. For both use cases, only the hidden layer is deployed in hardware simulation to capture energy metrics. The first layer's energy consumption is estimated based on its synaptic operations (SOPs) and the energy-per-SOP metric obtained from the second layer.

For the N-MNIST dataset, LOKI achieves a classification accuracy of \qty{98.0}{\%} with an estimated energy consumption of only \qty{119.8}{nJ} per inference. This highlights its capability to perform accurate and energy-efficient classification, making it ideal for systems integrating DVS sensors. Similarly, for the 1-word KWS task, LOKI effectively performs keyword spotting with an average performance of \qty{93.0}{\%} and an average inference energy of \qty{546.5}{nJ}, demonstrating its efficiency in processing spiking audio data. The accuracy results for each target keyword are summarized in \figref{fig:kws_results}{d}. These performances, combined with low energy consumptions, underscore the potential of LOKI to realize energy-efficient edge-AI applications.

\subsection{Comparison with prior work}

Table \ref{tab:comp} compares LOKI with state-of-the-art neuromorphic accelerators. By utilizing MCCG SRAM, LOKI achieves the highest clock frequency of \qty{667}{\mega\hertz} at \qty{0.59}{V}, with the exception of ESAM \cite{huijbregts2024}, which is using 3nm FinFET. LOKI has the lowest energy-per-SOP for FC SNN accelerators. Only SNE \cite{dimauro2022} achieves a slightly lower energy-per-SOP, however, it is targeted towards convolutional SNNs, which allows for weight sharing between synapses. Despite this, LOKI delivers state-of-the-art accuracy on both N-MNIST and KWS, while also having the lowest energy per inference for these tasks.

\section{Conclusion}
\label{sec:conclusion}
In this paper, we presented LOKI, an energy-efficient digital accelerator for FC SNNs. It implements Multi-Cycle Clock-Gated SRAM and block AER for achieving higher energy efficiency and throughput. Implemented in \qty{22}{\nano\meter} technology, LOKI only consumes \qty{0.266}{\pjsop}, the lowest of any state-of-the-art FC SNN accelerator. We evaluated LOKI on both the N-MNIST and KWS tasks. On N-MNIST, it achieves \qty{98.0}{\percent} accuracy while consuming \qty{119.8}{\njinf}, while on KWS, it achieves \qty{93.0}{\percent} accuracy while only consuming \qty{546.5}{\njinf}. These results showcase the potential of LOKI for integration with energy-efficient event-based sensors, enabling ultra-efficient edge-AI applications.

\section*{Acknowledgment}
This work is funded by EU’s Horizon Europe research and innovation programme under grant agreement No. 101070374.

\bibliographystyle{./IEEEtran.bst}
\bibliography{references}

\end{document}